\documentclass[twocolumn]{aastex63}
\usepackage{epsfig}
\usepackage{amsmath, amsthm, amssymb}
\usepackage{microtype}
\usepackage{float}
\usepackage{verbatim}
\usepackage{wrapfig}
\usepackage{graphicx}
\usepackage[caption = false]{subfig}
\usepackage{multirow}
\usepackage{hhline}
\usepackage{booktabs}
\usepackage{array}
\usepackage{bm}
\usepackage[normalem]{ulem}

\shorttitle{A unique candidate MSP}
\shortauthors{Swihart et al.}

\begin{document}

\title{4FGL J1120.0--2204: A Unique Gamma-ray Bright Neutron Star Binary with an Extremely Low Mass Proto-White Dwarf}

\correspondingauthor{Samuel J. Swihart}
\email{samuel.swihart.ctr@nrl.navy.mil}

\author{Samuel J. Swihart}
\affiliation{National Research Council Research Associate, National Academy of Sciences, Washington, DC 20001, USA,\\ resident at Naval Research Laboratory, Washington, DC 20375, USA}
\author{Jay Strader}
\affiliation{Center for Data Intensive and Time Domain Astronomy, Department of Physics and Astronomy,\\ Michigan State University, East Lansing, MI 48824, USA}
\author{Elias Aydi}
\affiliation{Center for Data Intensive and Time Domain Astronomy, Department of Physics and Astronomy,\\ Michigan State University, East Lansing, MI 48824, USA}
\author{Laura Chomiuk}
\affiliation{Center for Data Intensive and Time Domain Astronomy, Department of Physics and Astronomy,\\ Michigan State University, East Lansing, MI 48824, USA}
\author{Kristen C. Dage}
\affiliation{Department of Physics, McGill University, 3600 University Street, Montr\'{e}al, QC H3A 2T8, Canada}
\affiliation{McGill Space Institute, McGill University, 3550 University Street, Montr\'{e}al, QC H3A 2A7, Canada}
\author{Adam Kawash}
\affiliation{Center for Data Intensive and Time Domain Astronomy, Department of Physics and Astronomy,\\ Michigan State University, East Lansing, MI 48824, USA}
\author{Kirill V. Sokolovsky}
\affiliation{Center for Data Intensive and Time Domain Astronomy, Department of Physics and Astronomy,\\ Michigan State University, East Lansing, MI 48824, USA}
\author{Elizabeth C. Ferrara}
\affiliation{Department of Astronomy, University of Maryland, College Park, MD, 20742, USA}
\affiliation{Center for Exploration and Space Studies (CRESST), NASA/GSFC, Greenbelt, MD 20771, USA}
\affiliation{NASA Goddard Space Flight Center, Greenbelt, MD 20771, USA}

\begin{abstract}

We have discovered a new X-ray emitting compact binary that is the likely counterpart to the unassociated \emph{Fermi}-LAT GeV $\gamma$-ray source 4FGL J1120.0--2204, the second brightest \emph{Fermi} source that still remains formally unidentified. Using optical spectroscopy with the SOAR telescope, we have identified a warm ($T_{\textrm{eff}}\sim8500$ K) companion in a 15.1-hr orbit around an unseen primary, which is likely a yet-undiscovered millisecond pulsar. A precise \emph{Gaia} parallax shows the binary is nearby, at a distance of only $\sim 820$ pc. Unlike the typical ``spider" or white dwarf secondaries in short-period millisecond pulsar binaries, our observations suggest the $\sim 0.17\,M_{\odot}$ companion is in an intermediate stage, contracting on the way to becoming an extremely low-mass helium white dwarf (a ``pre-ELM" white dwarf). Although the companion is apparently unique among confirmed or candidate millisecond pulsar binaries, we use binary evolution models to show that in $\sim 2$ Gyr, the properties of the binary will match those of several 
millisecond pulsar--white dwarf binaries with very short ($< 1$ d) orbital periods. This makes 4FGL J1120.0--2204 the first system discovered in the penultimate phase of the millisecond pulsar recycling process. 

\vspace{10mm}
\end{abstract}

\section{Introduction}
Millisecond pulsars (MSPs) are neutron stars that have been spun-up (recycled) to very rapid spin periods via accretion from a binary companion \citep{Alpar82}. Despite the fact that this recycling process should be long-lived \citep{Tauris99}, most MSPs in the field of the Galaxy have completed the spin-up process, such that accretion has ceased, and the neutron star is left with a low-mass, helium white dwarf as a companion. This process is in contrast to MSPs in globular clusters, which likely acquired their current binary companion(s) via dynamical exchange interactions \citep[e.g.,][]{Hui10,Bahramian13,Ye19}. 

Nearly all MSPs emit $\gamma$-rays \citep{Abdo13}, and multiwavelength follow-up observations of previously unidentified \emph{Fermi}-LAT GeV $\gamma$-ray sources have resulted in an enormous increase in the number of known MSPs in the Galactic field \citep[e.g.,][]{Ray12}, particularly those in compact binaries that have non-degenerate, Hydrogen-rich companions. These binary systems are commonly referred to as ``spiders,'' due to the evaporative effects that the MSP has on its companion, and are further subdivided into ``black widows'' with ultra-light companions ($M_{c}\lesssim 0.05\,M_{\odot}$), and ``redbacks'' with more massive companions ($M_{c}\gtrsim 0.1\,M_{\odot}$) \citep{Roberts13,Strader19}.

Prior to the launch of \emph{Fermi}, most of these MSPs remained elusive: the energetic pulsar wind leads to substantial mass loss from the non-degenerate companion, eclipsing the radio emission and making it challenging to discover these systems in the short-timescale observations of all-sky radio surveys, especially at low observing frequencies \citep[e.g.,][]{Archibald13}. However, this radio eclipsing material does not affect the GeV emission, making follow-up of \emph{Fermi} sources a  successful method for identifying new MSP binaries with companions other than He white dwarfs.

One approach that has been used successfully to identify MSP counterparts in the unassociated \emph{Fermi} sources is to search for X-ray/optical matches inside the LAT error ellipses. X-rays in these binary systems are often produced by an intrabinary shock at the interface between the pulsar and companion winds \citep[e.g.,][]{Gentile14, Romani16, Wadiasingh17}, while the non-degenerate stellar companions typically display optical periodic variability due to tidal distortion of the secondary and/or irradiative heating of the tidally-locked companion surface \citep[e.g.,][]{McConnell15,Sanchez17,Cho18,Swihart18,Draghis19,Swihart19,Yap19,Swihart21}.

Although the number of spider MSPs has increased dramatically in recent years, binary evolution models predict these systems will not end up being the MSP--He white dwarf binaries that make up the bulk of the MSP binary population \citep[e.g.,][]{Benvenuto12,Chen13}. In contrast to redbacks in particular, systems with companions in the late stages of becoming a He white dwarf do not fill significant fractions of their Roche lobe since most of the companion envelope has either been accreted or ejected during MSP spin-up \citep[e.g.,][]{Podsiadlowski02,Postnov14,Ablimit19}. Such systems will not show substantial optical variability, and so have remained relatively elusive. 

The standard evolutionary pathway for creating MSP binaries with short-period, extremely-low-mass (ELM, $M\lesssim0.3\,M_{\odot}$) white dwarfs is well-understood \citep[e.g.,][]{Tauris99,Lin11,Istrate14,Istrate16}. However, there should be a population of systems in which the companion is still in an intermediate stage---after the outer envelope of the star has been lost during MSP spin-up, but before the companion fully contracts and reaches the white dwarf cooling branch \citep[e.g.,][]{Sarna01,Althaus13}.

Here we present the discovery of a new low-mass X-ray binary inside the bright, previously unassociated \emph{Fermi} $\gamma$-ray source 4FGL J1120.0--2204, and show it is likely a MSP with a warm companion in a relatively face-on orbit. Despite some similarities with the spider class of MSP binaries, the companion is more consistent with being a precursor to an extremely-low-mass white dwarf, commonly referred to as a ``pre-ELM'' white dwarf \citep[e.g.,][]{Maxted11,Rappaport15,Pelisoli19}. This places the system in an intermediate stage on its way to becoming a short-period MSP--He white dwarf binary, and so is likely in the late phases of the standard evolutionary pathway leading to the bulk of the MSP population.

\begin{figure*}[t!]
\begin{center}
  \subfloat{\includegraphics[width=0.46\textwidth]{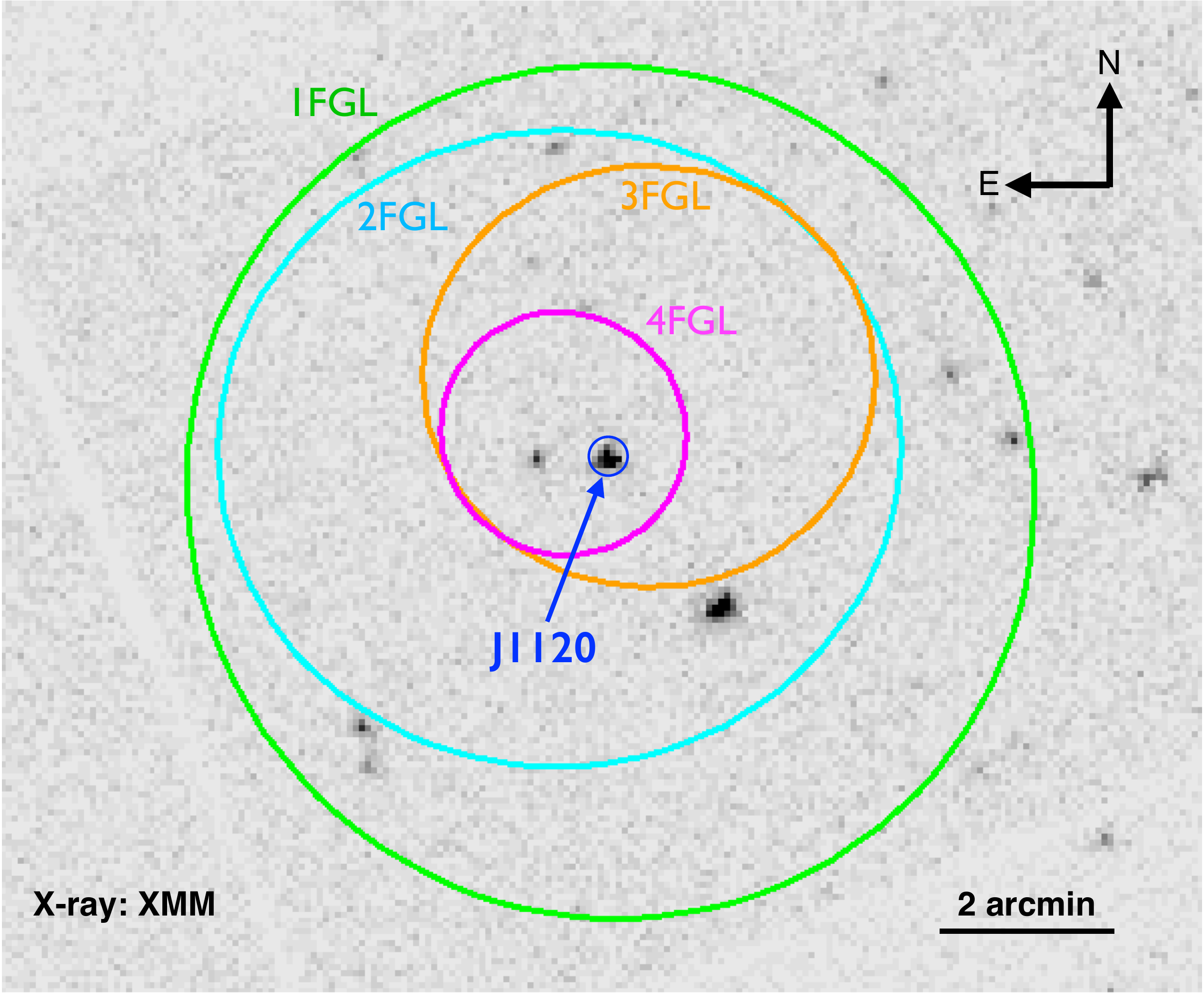}}
  \subfloat{\includegraphics[width=0.5\textwidth]{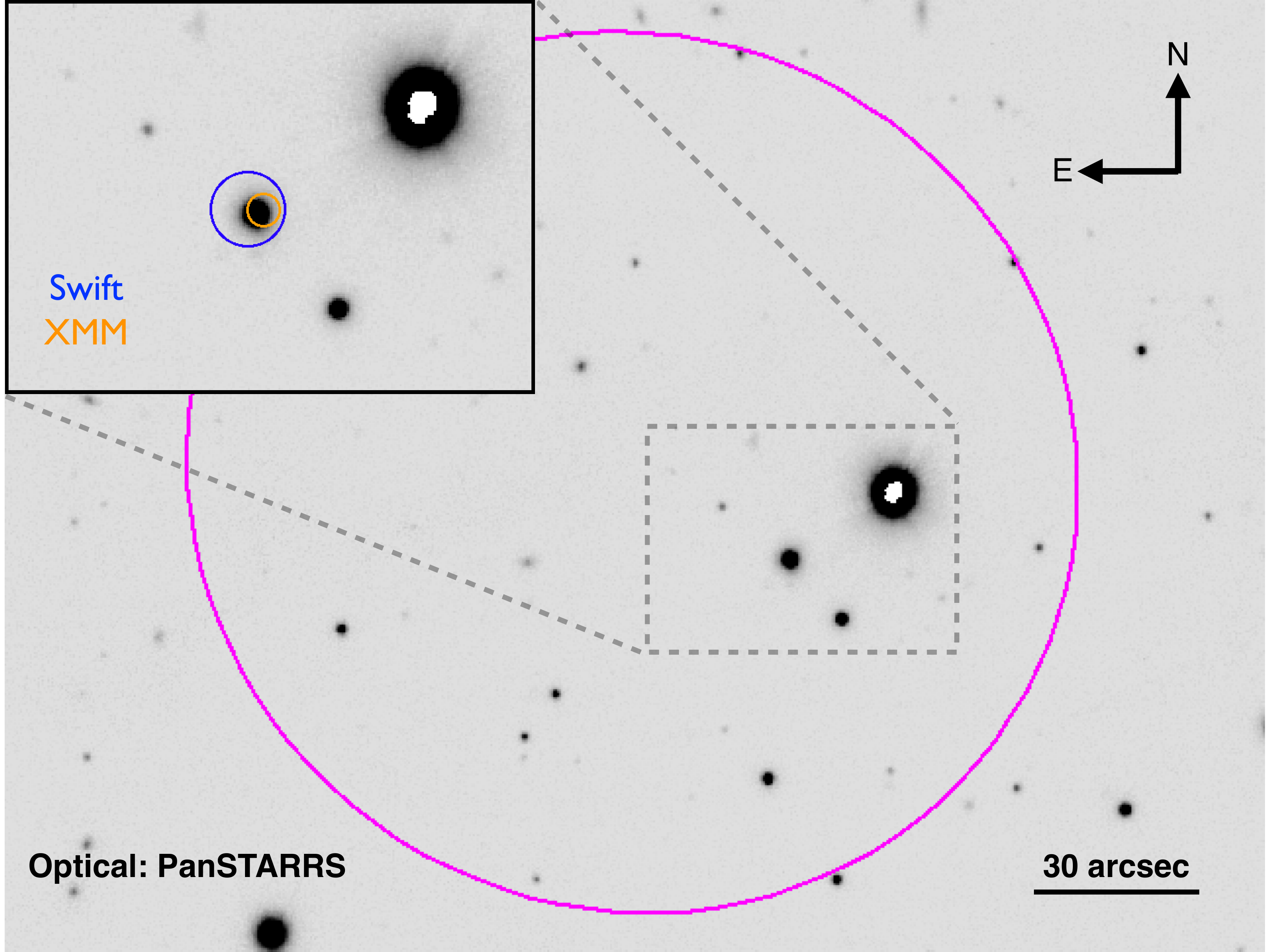}}
\caption{Left: \emph{XMM-Newton}/EPIC X-ray image of the field showing the positions and overlapping 95\% error ellipses from the 1FGL, 2FGL, and 3FGL catalogs corresponding to the $\gamma$-ray source 4FGL J1120.0--2204 (magenta). The likely counterpart to the $\gamma$-ray source that is the subject of this work is marked in blue. The fainter X-ray source visible in the 4FGL ellipse is a galaxy unrelated to the $\gamma$-ray emission \citep{Salvetti17}. Right: Optical Pan-STARRS $i'$ image zoomed-in on the 4FGL region. The inset displays a further zoomed-in view showing the positions (90\% confidence) of the \emph{Swift} and \emph{XMM} source marked in the left panel. The optical source coincident with the X-rays (hereafter, J1120) is the likely companion in a MSP binary (see Sec.~\ref{sec:results}).}
\label{fig:finders}
\end{center}
\end{figure*}

\section{Observations}

\subsection{The $\gamma$-ray Source}
\label{sec:gammarays}
The unassociated \emph{Fermi} source 4FGL J1120.0--2204 is a bright, persistent $\gamma$-ray source that has been known since it was first catalogued as 1FGL J1119.9--2205 based on the first eleven months of survey data. Since then, it has appeared in each subsequent LAT catalog (2FGL J1120.0--2204, 3FGL J1119.9--2204), and is currently listed in the 4FGL-DR2 catalog \citep{Ballet20} with a detection significance of 54.9$\sigma$, making it the second brightest \emph{Fermi} source that still remains formally unidentified (the brightest is 4FGL J2112.5--3043, which has a detection significance of 56.3$\sigma$). The updated 4FGL 95\% confidence error ellipse lies within the 3FGL region and is 69\% smaller in area (Figure~\ref{fig:finders}). The $\gamma$-ray spectrum is highly curved, with a curvature significance of 16.0$\sigma$ for a LogParabola model (16.3$\sigma$ for a cut-off power law), a common feature among \emph{Fermi}-detected MSPs \citep{Abdo10}. In fact, a number of independent groups have classified this \emph{Fermi} source as a highly-likely MSP based on machine learning classifications using the $\gamma$-ray data \citep{Saz16,Kaur19,Finke20,Kerby21,Bhat21}.

The 0.1--100 GeV flux, $F_{\gamma} = (1.69 \pm 0.06) \times 10^{-11}$ erg cm$^{\textrm{-2}}$ s$^{\textrm{-1}}$, corresponds to a luminosity of $(1.35 \pm 0.05) \times 10^{33}\,(d/0.82\,\textrm{kpc})^{\textrm{2}}$ erg s$^{\textrm{-1}}$, assuming the geometric distance estimate from a significant \emph{Gaia} parallax measurement of the suspected optical counterpart (Sec.~\ref{sec:dist}).

\subsection{X-rays}
\label{sec:X-rays}

\subsubsection{Swift}
\label{sec:Swift}
The field containing 4FGL J1120.0--2204 was observed with \emph{Swift}/XRT, targeting the 1FGL and 2FGL regions on 27 epochs between 2010 Nov and 2013 Mar, for a total of $\sim$65 ksec. After the release of the 3FGL catalog, \citet{Hui15} first reported on the X-ray sources inside the overlapping 2FGL and 3FGL regions. Their analysis found two X-ray sources detected at $>4\sigma$ inside the 3FGL error ellipse. These sources also lie within the 4FGL region. The brightest of these sources (listed as J1120\_X1 in \citealt{Hui15}) is the subject of this work and, as we argue below, is the likely MSP counterpart to the $\gamma$-ray source. Hereafter, we refer to this X-ray source (and the corresponding optical source, see below) as J1120.

\citet{Hui15} calculated the X-ray flux of this source by assuming the Galactic HI column density ($N_{H} = 3.9\times10^{20}\,\textrm{cm}^{\textrm{-2}}$) \citep{Kalberla05} and a power-law model with $\Gamma = 2$, finding an unabsorbed 0.3--10 keV flux of $(4.7 \pm 0.5) \times 10^{-14}$ erg cm$^{\textrm{-2}}$ s$^{\textrm{-1}}$. No spectral or temporal analysis of these data was carried out by this group.

\citet{Kerby21} performed a more detailed spectral analysis of this source, finding a best fit power law index of $\Gamma = 1.89^{+0.18}_{-0.17}$ and unabsorbed 0.3--10 keV flux of $(7.9^{+1.2}_{-1.0}) \times 10^{-14}$ erg cm$^{\textrm{-2}}$ s$^{\textrm{-1}}$. It is notable that despite using the same datasets, these two flux estimates do not agree within the uncertainties. Even if we account for the slightly different spectral models assumed, we find no straightforward explanation for this discrepancy, though it does not substantively affect any of our conclusions.

For the remainder of this work, we assume the X-ray properties derived from the more recent \emph{XMM} data presented below.

\subsubsection{XMM}
\label{sec:XMM}
This region was also observed by \emph{XMM-Newton} for $\sim$74 ksec in 2014 June. Prior to the release of the 4FGL catalog, \citet{Salvetti17} analyzed data from all sources inside and around the 3FGL region, focusing specifically on sources that had optical counterparts spatially coincident with the X-ray source positions. Three of the X-ray sources they analyzed had plausible optical counterparts, and two of these are present within the 4FGL 95\% confidence ellipse (see Figure~\ref{fig:finders}; we note these are the same two sources identified by \citet{Hui15} from the \emph{Swift} data). As pointed out by \citet{Salvetti17}, the fainter of these two X-ray sources is a known galaxy, and is unlikely to be the source of the $\gamma$-ray emission, especially given the highly curved $\gamma$-ray spectrum; \citet{Kerby21} estimate the probability that this 4FGL source is a blazar is only $\sim$3\%, supporting the inference that the fainter X-ray source is unrelated to the $\gamma$-ray emission.

The brightest \emph{XMM} source in the 4FGL ellipse (Figure~\ref{fig:finders}, left panel, blue circle) is spatially coincident with an optical source whose properties are consistent with a stellar companion to a yet-undiscovered MSP (see Sec.~\ref{sec:results}). \citet{Salvetti17} analyzed the \emph{XMM} X-ray spectrum of this source (assuming a Galactic column density of $N_{H} = 4 \times 10^{20}$ cm$^{-2}$), finding $\Gamma = 2.63^{+0.12}_{-0.11}$ and an unabsorbed 0.3--10 keV flux of $(7.3^{+0.6}_{-0.5}) \times 10^{-14}$ erg cm$^{\textrm{-2}}$ s$^{\textrm{-1}}$ (68\% confidence), equivalent to a 0.5--10 keV X-ray luminosity of $(4.1\pm0.3) \times 10^{30}$ erg s$^{-1}$ at the \emph{Gaia} distance of the source.

While the X-ray fluxes of the \emph{XMM} and \emph{Swift} observations match within their uncertainties, the photon indices do not; formally the \emph{XMM} spectral fit is softer at the $>3\sigma$ level. Either value is within the range of photon indices observed for known MSP binaries at this X-ray luminosity 
\citep{Lee18}. One possibility is that the uncertainty in one or both of the measurements is underestimated; another is that the X-ray spectrum has changed between the two observations. In close MSP binaries where the X-rays originate from an intrabinary shock at the interface between the pulsar and companion winds, the X-rays can vary on both orbital and secular timescales (e.g., \citealt{AlNoori18}). However, the X-ray fluxes of the two observations are consistent with each other, and \citet{Salvetti17} found no evidence for variability in the 
\emph{XMM} light curve. We also checked to see whether the GeV flux and spectrum changed between the \emph{Swift} and \emph{XMM} observations, but found no evidence for a change in either quantity, and we find no significant variability in the long-term $\gamma$-ray light curve. It is therefore likely safe to assume that the X-ray spectral index change between the \emph{Swift} and \emph{XMM} observations is not associated with a genuine state-transition as is observed in the known transitional MSPs \citep[e.g.,][]{Torres17}.

We folded the \emph{XMM} data on the orbital period derived in Sec.~\ref{sec:orb_params} and confirm the source shows no orbital variability, which may be due to a combination of a weak expected intrabinary shock and the relatively face-on orbital geometry of the system  (Sec.~\ref{sec:results}). Future X-ray observations would be useful to further explore the X-ray spectrum and variability
of J1120.

\subsection{Optical Counterpart}
\label{sec:opt_counterpart}
The brightest X-ray source in the 4FGL region clearly matches to a bright ($G=15.5$) and relatively blue ($BP - RP = 0.24$) source in \emph{Gaia} EDR3 \citep{Gaia21}, with a ICRS position of (R.A., Dec.) = (11:19:58.309, --22:04:56.33), which we take as the best position available. No other optical sources are present within $14\arcsec$ (down to a limiting magnitude of $i'\sim22.5$ mag).  This source has a significant EDR3 parallax, 1.18 $\pm$ 0.04 mas, corresponding to a precise geometric distance of $816.8^{+22.0}_{-24.8}$ pc \citep{BailerJones21} after a parallax zeropoint correction \citep{Lindegren21}, and as discussed below (Sec.~\ref{sec:dist}), has a large proper motion due to its high space velocity. 

This optical source (J1120) is also present in the Pan-STARRS DR2 \citep{Chambers16} and Skymapper DR2 \citep{Skymapper19} surveys. The right panel of Figure~\ref{fig:finders} shows a Pan-STARRS $i'$ image of the field corresponding to 4FGL J1120.0--2204 with the positions (90\% confidence) of the \emph{Swift} and \emph{XMM} X-ray sources overlaid in the inset.

As discussed in \citet{Hui15} and \citet{Salvetti17}, J1120 is also detected in Catalina Sky Survey data \citep[CSS;][]{Drake09} and by the Gamma-Ray Burst Optical/Near-Infrared Detector \citep[GROND,][]{GROND08}. Both groups attempted to find periodic variations in these photometry but were unsuccessful. Previewing our conclusions about this source, we suspect no variations are present due to the relatively face-on orientation of the binary orbit, the secondary underfilling its Roche lobe, and negligible current heating of the companion.

This source is also present in the All-Sky Automated Survey for Supernovae \citep[ASAS--SN,][]{Shappee14,ASASSN17}, with a brightness of $g'=14.42$ mag. However, the ASAS--SN pixels are large ($\sim8\arcsec$), so this measurement may be confused by the bright ($V=12.4$) source that is only $14.5\arcsec$ away (the saturated star in the right panel of Figure~\ref{fig:finders}). Similar to the CSS and GROND datasets, no periodic variability is present in the ASAS-SN photometry.

\subsection{Optical Spectroscopy}
Motivated by the spatial coincidence between the X-ray and optical source, we obtained 40 usable spectra of J1120 with the Goodman Spectrograph \citep{Clemens04} on the 4.1-m SOAR telescope between 2019 Apr and 2021 Jul. Nearly all the data were obtained in 2021.

The spectra were obtained in three main configurations, depending on the observing conditions and grating availability: a 1200 l mm$^{-1}$ grating with a resolution of 1.7 \AA\ FWHM and wavelength coverage $\sim 5500$--6750 \AA; a 2100 l mm$^{-1}$ grating with a resolution of 0.8 \AA\ FWHM and wavelength coverage $\sim 5630$--6230 \AA; and the same 2100 l mm$^{-1}$ grating, instead covering $\sim 6020$--6610 \AA. The spectra were taken through a 0.95\arcsec\ longslit. Typical exposure times were 1200s per spectrum, though a few were slightly shorter or longer.

The data were reduced and optimally extracted in the standard manner using tools in IRAF \citep{Tody86}. We derived radial velocities through cross-correlation with the highest signal-to-noise spectrum of J1120 obtained in each setup. These velocities are listed in Table~\ref{tab:soar_rv_tab}.

\section{Results and Analysis}
\label{sec:results}

\subsection{Stellar Parameters and Unusual Abundances}
\label{sec:spec_results}

\begin{figure}[t!]
\includegraphics[width=\linewidth]{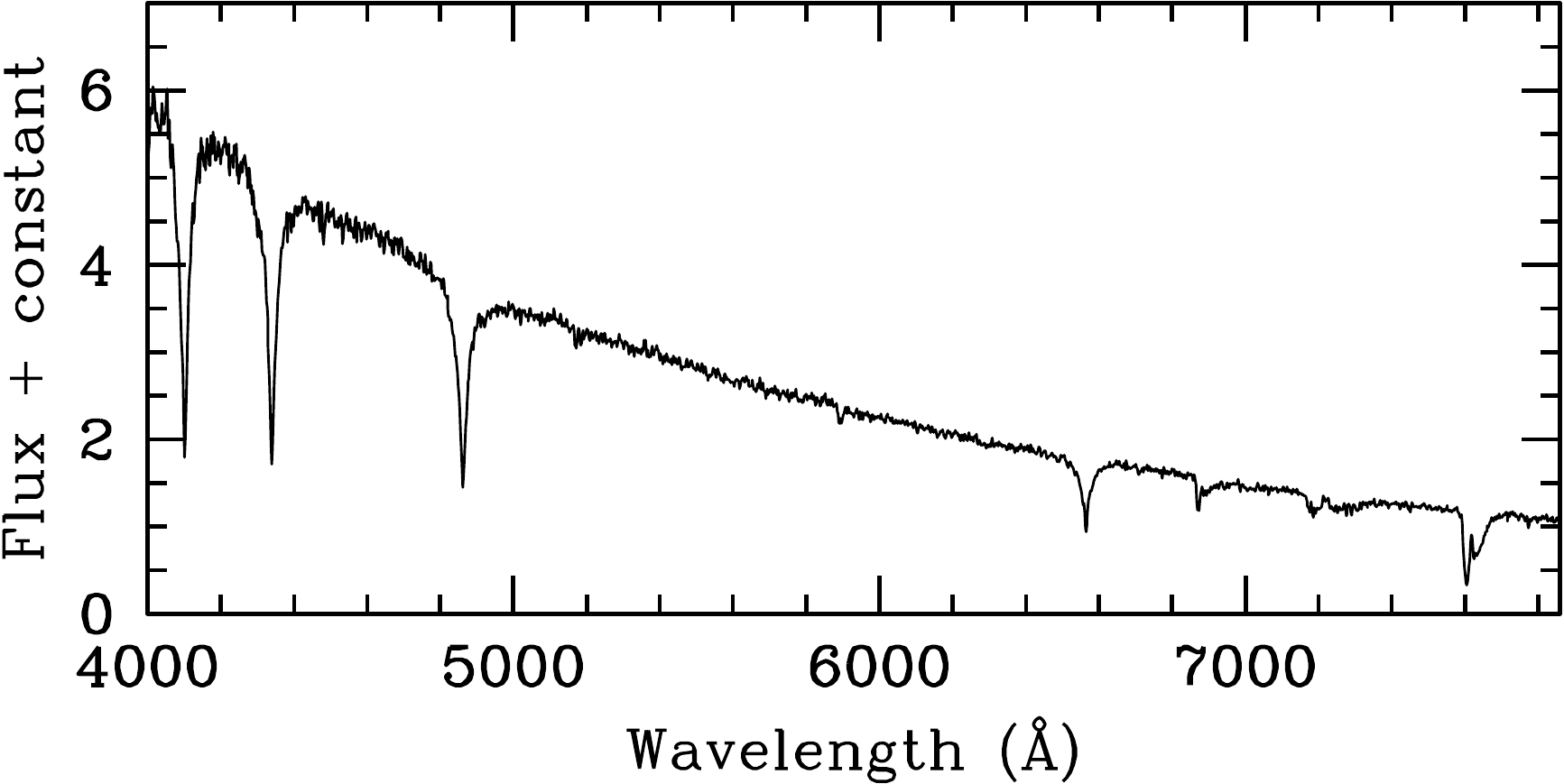}
\caption{Low-resolution SOAR spectrum of J1120 with a relative flux calibration applied. The star has a blue continuum, and the strongest intrinsic features visible are broad Balmer lines. Weak Mg and Na absorption is also apparent.}
\label{fig:j1120_spec_lr}
\end{figure}

The single low-resolution flux-calibrated spectrum of J1120 has a blue continuum dominated by broad Balmer absorption lines (Figure~\ref{fig:j1120_spec_lr}), strongly suggesting an early to mid-A spectral type. No \ion{He}{1} lines are visible. Other moderately strong absorption lines include Ca K (the Ca H line is buried in H$\epsilon$), the Mg triplet, and what appears to be \ion{O}{1} at 7773 \AA. Absorption at the Na doublet is also present, which is a blend of interstellar and stellar absorption (see below). We do not see any emission lines.

\begin{figure}[t!]
\includegraphics[width=\linewidth]{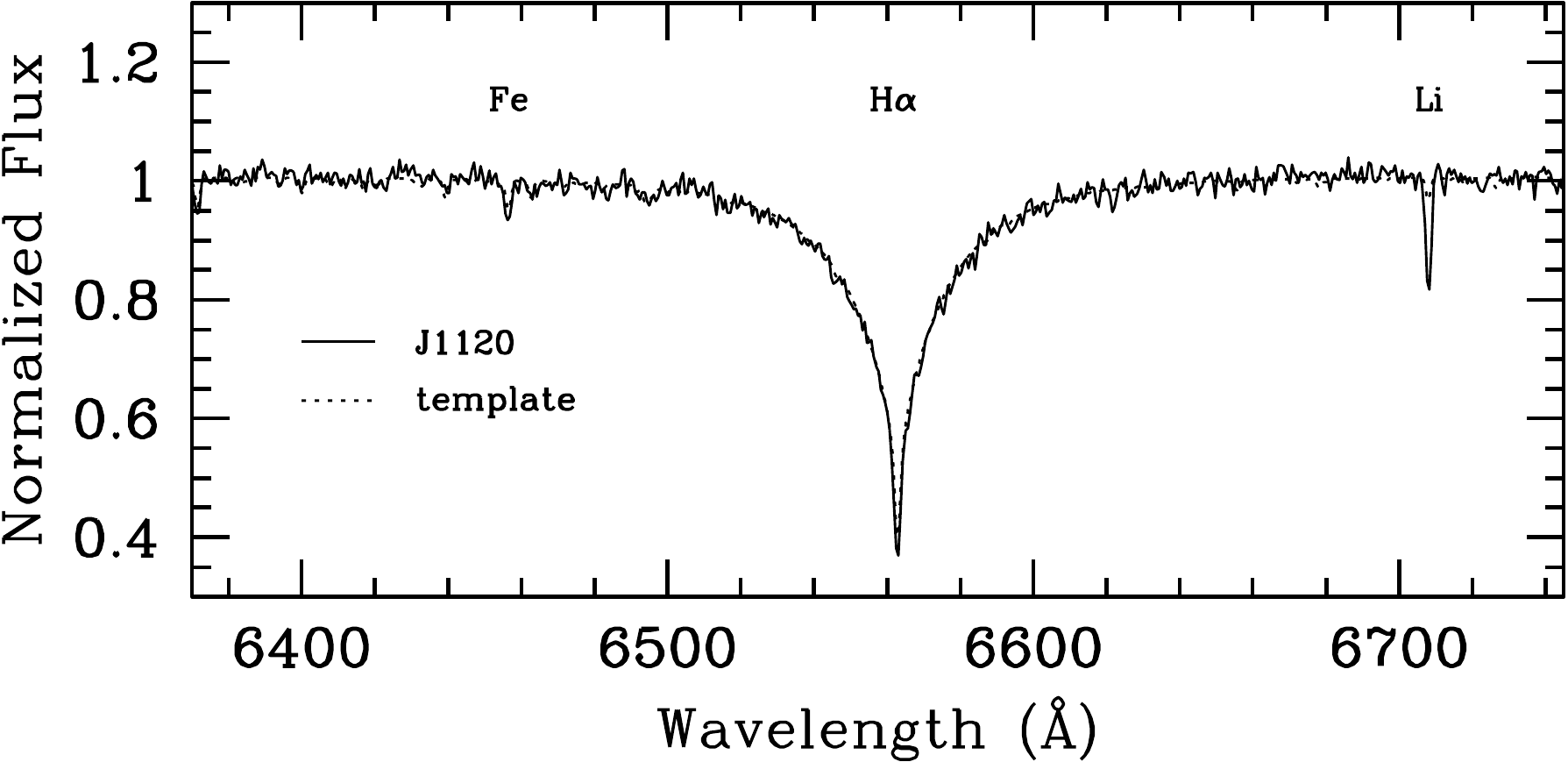}
\caption{Continuum-normalized medium-resolution spectrum of J1120 (solid line) from 2021 Feb 19 in the region of H$\alpha$. A resolution-matched solar metallicity PHOENIX template with $T_{\rm eff} = 8600$ K and log $g = 4.5$ cm s$^{-2}$ is overplotted (dashed line). The \ion{Li}{1} line at 6708 \AA\ in the J1120 spectrum is very strong.}
\label{fig:j1120_spec_hr}
\end{figure}

To help constrain the effective temperature and gravity of the star, we fit the highest signal-to-noise 1200 l mm$^{-1}$ spectra to a grid of solar metallicity PHOENIX models \citep{Husser13} in the region of the H$\alpha$ line that were smoothed to the resolution of the SOAR data. Given the stellar radius and binary inclination inferred below, the projected rotation of the star can be neglected at our spectral resolution. We find best-fit values of $T_{\rm eff} = 8650\pm110$ K and log $g = 4.6\pm0.2$ cm s$^{-2}$ ($\chi^{2}/\textrm{dof} = 212.8/211$). Since J1120 likely has an unusual structure and abundances (see below), these values could have substantial systematic uncertainties, and we take them more as general guidance than precise measurements. In Figure~\ref{fig:j1120_spec_hr}, we overplot the closest template to these best-fit values ($T_{\rm eff} = 8600$ K; log $g = 4.5$ cm s$^{-2}$) on one of the medium-resolution J1120 spectra from 2021 Feb 19 in the region of the H$\alpha$ line.

While these solar metallicity models do a reasonable job of matching the H$\alpha$ region, they poorly fit most of the metal lines. We also experimented with fits of different bulk metallicity ($Z$), finding that these were much worse fits than solar metallicity. This indicates that the poor metal-line fits are due to unusual abundances rather than solar-scaled abundances at a lower or higher metallicity.

One clear example of this is for \ion{Na}{1}. In the two high signal-to-noise 2100 l mm$^{-1}$ spectra covering the Na doublet, there are two clear doubled-lined components. The weaker component (equivalent width 0.26 \AA~summed across both lines), with a velocity near rest, is consistent with interstellar absorption. The stronger component (summed equivalent width $0.51\pm0.02$ \AA) has a velocity consistent with the other absorption lines and hence is associated with J1120. These lines are stronger than predicted for a star with solar abundances and the inferred $T_{\rm eff}$ and log $g$, which from the appropriate PHOENIX models is an equivalent width of $\sim 0.20$ \AA. Even for somewhat cooler temperatures (i.e., $T_{\rm eff} \sim 8000$ K, as inferred from the photometry presented below), the observed \ion{Na}{1} line strength is significantly larger than predicted. This hints at an overabundance of Na in J1120.

The other unusual line apparent in the 1200 l mm$^{-1}$ spectra is \ion{Li}{1} at 6708 \AA~(Figure~\ref{fig:j1120_spec_hr}). It is among the strongest metal lines visible in these spectra (equivalent width $0.34\pm0.03$ \AA), but is predicted to be quite weak (equivalent width 0.04 \AA). As for Na, this comparison again suggests a substantial overabunance of Li. 

It should be noted that this is not the only suspected MSP binary that shows a substantial overabundance of Li. In the stripped redback PSR J1740--5340 in the globular cluster NGC 6397, a very high Li content is observed in a series of high-resolution spectra \citep{Sabbi03}. Those authors suggest the most plausible explanation for this overabundance is that fresh $^7$Li is being produced on the stellar surface due to bombardment from the high-energy pulsar wind. Since we currently observe minimal heating on the companion surface, it is difficult to explain the high Li abundance with this scenario. However, heating from the suspected neutron star may have been much more significant in the past, and large amounts of Li may still be present from these times. Additional explanations are also possible, such as a recent shell flash in the proto-white dwarf that could have produced some of the Li. Further study will be needed in order to properly interpret the high Li abundance.

The secondaries of other low-mass X-ray binaries have also shown Li overabundances (e.g., Cen X-4; \citealt{Gonz05}). For the purpose of this paper, the main implication of the Li measurement is that it proves that the companion star is definitely unusual and provides additional evidence that J1120 is indeed the counterpart to the GeV $\gamma$-ray source.

Given the modest resolution of our data and the non-standard structure of the star, we do not attempt to convert the above equivalent widths into abundances. Due to the brightness of the star ($G=15.5$), it should be feasible to obtain a high-resolution spectrum of sufficient quality to allow a proper abundance analysis in the future.

\begin{figure}[t!]
\begin{center}
	\includegraphics[width=\linewidth]{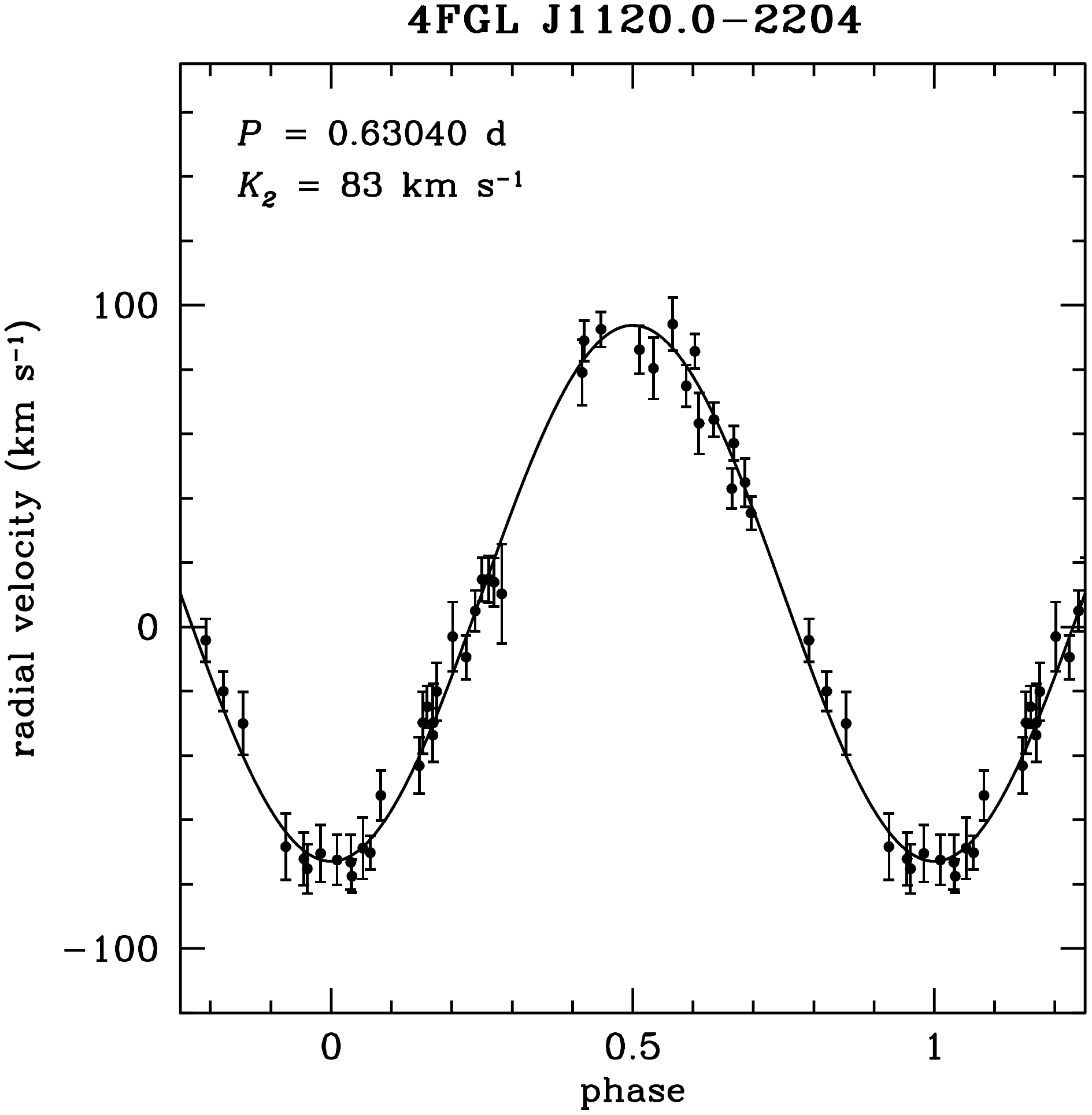}
    \caption{Circular Keplerian fit to the SOAR/Goodman barycentric radial velocities of the optical counterpart to 4FGL J1120.0--2204.}
\label{fig:rv_fig}
\end{center}
\end{figure}

\subsection{Orbital Paramters}
\label{sec:orb_params}

We fit circular Keplerian models to the radial velocities using the Markov Chain Monte Carlo sampler \texttt{TheJoker} \citep{PriceWhelan17}, fitting for the binary period $P$, BJD time of ascending node $T_{0}$, orbital semi-amplitude $K_{2}$, and systemic velocity $\gamma$. We find $P = 0.630398(27)$ d, $T_0 = 2459307.116 \pm 0.031$ d, $K_{2} = 83.3 \pm 1.7$ km s$^{-1}$, and $\gamma = 10.4\pm1.3$ km s$^{-1}$. Uncertainties are given at 1$\sigma$. We show this best-fit model along with the data in Figure~\ref{fig:rv_fig}. This is a good fit with $\chi^2$/d.o.f. = 30/36 and an r.m.s. scatter of 6.8 km s$^{-1}$. We also experimented with eccentric fits, finding an eccentricity posterior peaked at $e=0$ and no improvement in the reduced $\chi^2$. Hence we see no evidence for a non-zero eccentricity.

We use the posterior samples from the circular fit to derive the binary
mass function $f(M)$,

\begin{equation}
f(M) = \frac{P K_{2}^{3}}{2 \pi G} = \frac{M_1 \, (\textrm{sin}\, i)^{3}}{(1+q)^{2}}
\end{equation}

for mass ratio $q=M_{2}/M_{1}$ and inclination $i$, finding $f(M) = 0.0377 \pm 0.0023\, M_{\odot}$. Since this value represents the lower limit on the mass of the primary object for an edge-on (90$^{\circ}$) orbit, these results immediately suggest the orbit must be relatively face-on in order to accommodate a neutron star-mass primary.

\begin{figure}[t!]
\begin{center}
	\includegraphics[width=\linewidth]{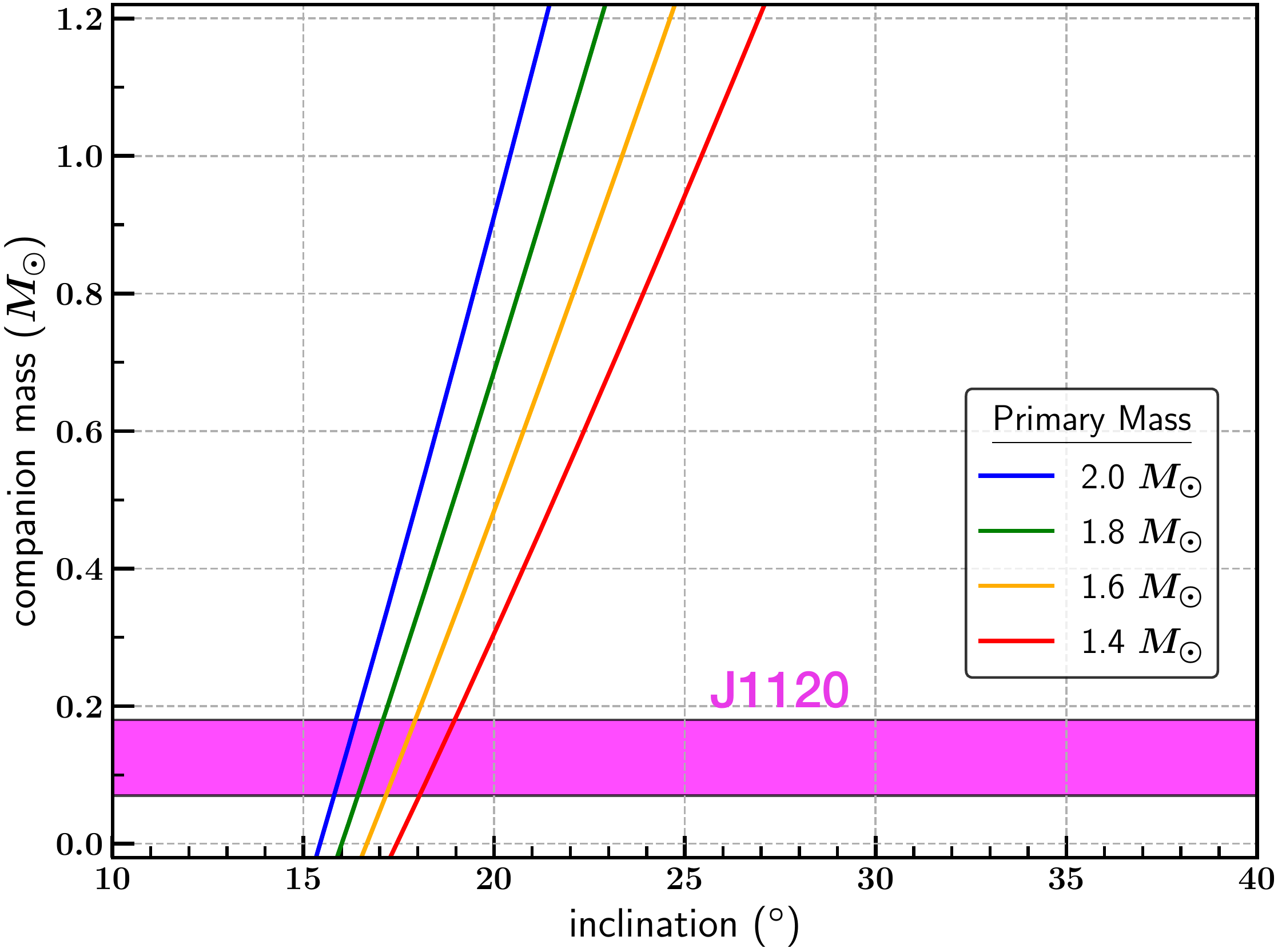}
    \caption{The range of possible inclinations and companion masses that can accommodate a neutron star-mass primary (solid lines), given our spectroscopic results. The shaded region represents the most likely mass of the secondary based on the system parameters and our binary evolution models (see below), which suggest the system is on its way to becoming a low-mass white dwarf ($M\sim0.17\,M_{\odot}$), and implying an inclination of $i \sim 16^{\circ}$--$19^{\circ}$ for typical neutron star masses.}
\label{fig:incl}
\end{center}
\end{figure}

Figure~\ref{fig:incl} shows the range of inclinations and companion masses that are possible given the spectroscopic results for a range of neutron star masses (solid lines). The shaded region represents the most likely mass of the secondary based on the system parameters and binary evolution models of the system (see below), which suggest the system is on its way to becoming a low-mass white dwarf ($M\sim0.17\,M_{\odot}$), implying an inclination of $i \sim 16^{\circ}$-- $19^{\circ}$ for a neutron star in the mass range $\sim 1.4-2.0\,M_{\odot}$. This very face-on inclination is the likely explanation for the lack of obvious orbital variability
(see Sec.~\ref{sec:comp_size}).

\subsection{Properties of the Companion}
\label{sec:companion}

\subsubsection{Colors \& Temperature}
\label{sec:temps}
\begin{figure}[t]
\begin{center}
	\includegraphics[width=1.0\linewidth]{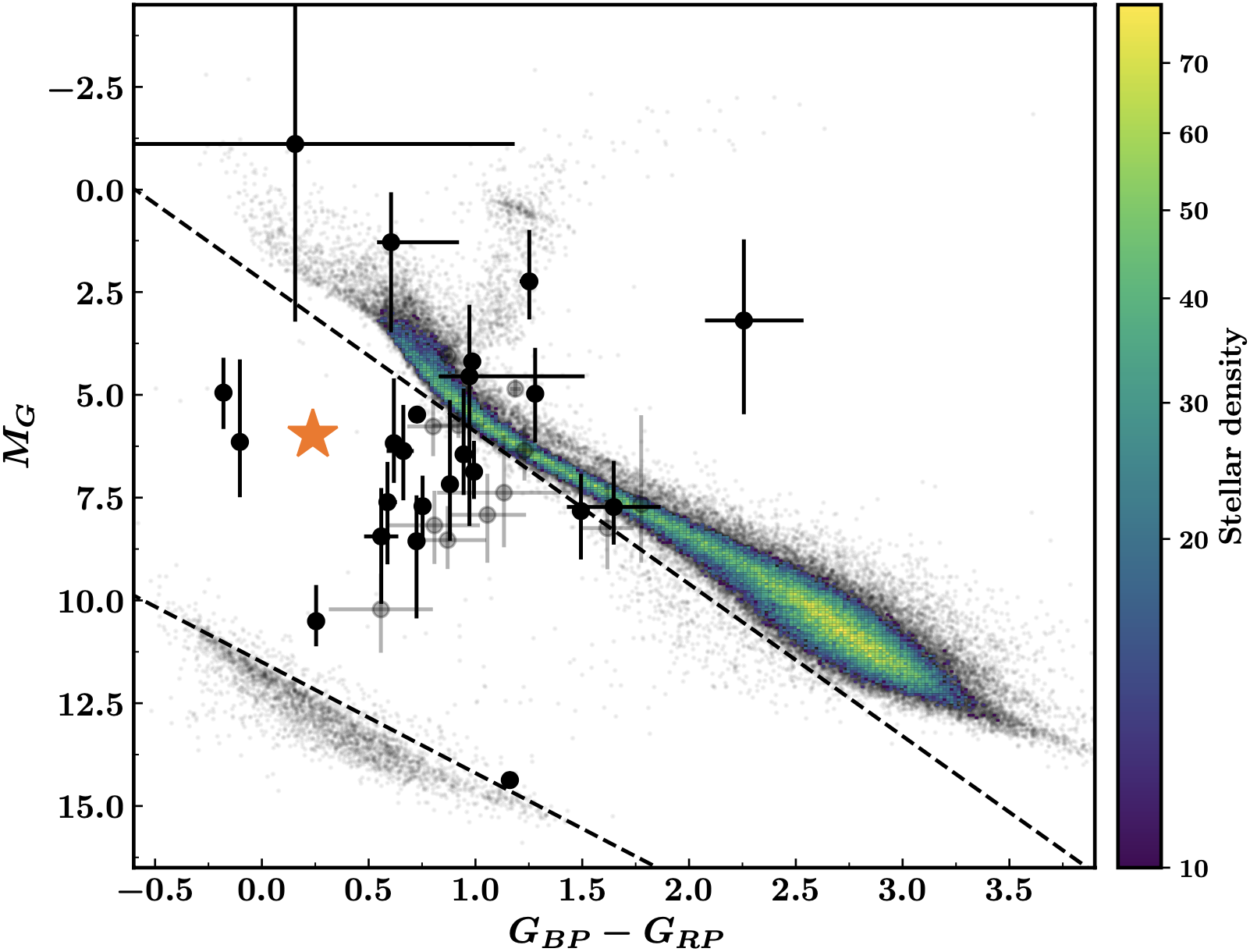}
    \caption{Observational color--absolute magnitude diagram (data from \emph{Gaia} EDR3) of field stars with precise parallax measurements within 200 pc from the Sun (scatter points \& colored histogram). The positions of confirmed and candidate MSP companions are shown with black and grey circles, respectively. The black dashed lines are empirical relations that filter 99.8\% of the field stars and white dwarfs in this figure. Most of the MSP companions are bluer and brighter than main-sequence stars of similar masses, suggesting they are highly irradiated or have been stripped of their outer envelopes. The suspected optical counterpart to 4FGL J1120.0--2204 (gold star) is much warmer than main-sequence stars at a similar brightness and is significantly brighter than white dwarfs with comparable colors. The two sources blueward of J1120 and with similar $M_{G}$ are PSR J1816+4510 \citep{Kaplan12,Kaplan13} (see Sec.~\ref{sec:endfate}) and PSR J0337+1715, a MSP in a triple system with two low-mass white dwarfs \citep{Ransom14}. Figure adapted from \citet{Antoniadis21}.
    }
\label{fig:cmd_fig}
\end{center}
\end{figure}

As mentioned in Sec.~\ref{sec:opt_counterpart}, J1120 has a blue color in \emph{Gaia}. This is confirmed in the Pan-STARRS and Skymapper data: the dereddened $(g' - i')_0$ colors are --0.28 and --0.32, respectively, where band-dependent extinction corrections were applied using the \citet{Schlafly11} reddening maps. In fact, this source lies in a peculiar region of the color-magnitude diagram (CMD), as shown in Figure~\ref{fig:cmd_fig}. This figure, which has been adapted from \citet{Antoniadis21}, shows an observational color--absolute magnitude diagram of field stars with precise parallax measurements within 200 pc from the Sun (scatter points), along with the positions of the known (black circles) and candidate (grey circles) MSP companions that are listed in \emph{Gaia} EDR3. Most of these MSP companions are bluer and brighter than main-sequence stars of similar masses, suggesting they are highly irradiated or have been stripped of their outer envelopes \citep{Antoniadis21}. The black dashed lines are empirical relations that filter 99.8\% of the field stars and white dwarfs in this figure.

J1120, the suspected optical counterpart to 4FGL J1120.0--2204, clearly lies in a sparsely populated region on this diagram (Figure~\ref{fig:cmd_fig}, gold star), with a blue color suggestive of a warm atmosphere compared to main sequence stars, but with a luminosity significantly brighter than white dwarfs with comparable colors.

The temperature of J1120 was estimated by \citet{Bai19} using a machine learning algorithm to predict the effective temperatures of $\sim$133 million stars in \emph{Gaia} DR2 using their proper motions and ($G - G_{\textrm{RP}}$) and ($G_{\textrm{BP}} - G$) colors. Using this technique, they estimate $T_{\textrm{eff}} = 8001 \pm 304$ K. We note that given the unusual abundances presented above, this estimate should not be taken as a precise measurement.

As an independent estimate of the temperature inferred from the photometry, we compare the de-reddened Pan-STARRS and Skymapper colors to the theoretical colors of main sequence stars using a 100 Myr solar metallicity isochrone \citep{Bressan12}, similar to the procedure described in \citet{Swihart20}. We find $T_{\textrm{eff}} = 8430$ K and $T_{\textrm{eff}} = 8530$ K for the Pan-STARRS and Skymapper colors, respectively. Assuming a more metal-poor star with [Fe/H] = –1 gives consistent results within $\sim$100 K. These values are broadly consistent with our estimate from the spectral modeling (Sec.~\ref{sec:spec_results}). Overall, given our analysis of the spectra and colors, for the remainder of the paper we assume $T_{\rm eff} = 8500\pm200$ K.

\subsubsection{Distance and Kinematics}
\label{sec:dist}
J1120 has a significant Gaia EDR3 \citep{Gaia21} parallax, corresponding to a geometric distance of $d = 816.8^{+22.0}_{-24.8}$ pc \citep{BailerJones21}. It also has a large proper motion ($36.33\pm0.02$ mas yr$^{\textrm{-1}}$), corresponding to a tangential space velocity of $140.6^{+3.9}_{-4.3}$ km s$^{\textrm{-1}}$. High space velocities are expected in binary MSPs \citep[e.g.,][]{Hobbs05,Desvignes16,Lynch18}, providing additional supporting evidence for the MSP nature of this system. If our MSP interpretation is correct, this large space velocity is likely the signature of a natal kick from the supernova that birthed the suspected neutron star primary.

The Galactic coordinates of J1120 are ($l\sim276.5^{\circ}$, $b\sim36.0^{\circ}$), which corresponds to a height of $\sim$480 pc above the midplane of the Galactic disk given its distance $d$. Assuming the systemic radial velocity from Sec.~\ref{sec:orb_params} together with the distance and proper motion measurements from \emph{Gaia} (and their 1$\sigma$ uncertainties), we derive the 3-D space motion of the system corrected to the local standard of rest \citep{Coskunoglu11}, finding $(U,V,W)_{\rm{LSR}} = (130 \pm 4, -14.2 \pm 0.6, 8.0 \pm 0.4)$ km s$^\textrm{-1}$. The corresponding 3-D space velocity ($131 \pm 4$ km s$^\textrm{-1}$) is consistent with the velocities of binary MSPs \citep[e.g.,][]{Gonzalez11,Matthews16}. Another explanation for the high velocity could be that the star is a member of the Galactic halo. However, typical halo stars in the solar neighborhood ($d\lesssim3$\,kpc) have 3-D space velocities $\gtrsim$220 km s$^{\rm{-1}}$ \citep[e.g.,][]{Nissen10,Bonaca17}, so it is unlikely J1120 is a member of the halo.

\subsubsection{Companion Size \& Lack of Variability}
\label{sec:comp_size}
The lack of optical variability in J1120 suggests the system is relatively face-on, which is supported by our spectroscopic results, but also is consistent with the secondary only partially filling its Roche lobe. We test this hypothesis by estimating the size of the companion and comparing it to the size of its full Roche lobe.

Assuming the \emph{Gaia} $G$ mag of the secondary, including extinction from the \citet{Green19} dustmaps, and a distance of 817 pc, the absolute magnitude is $M_{G} = 5.85$. Assuming $T_{\textrm{eff}} = 8500 \pm 200$ K (Sec.~\ref{sec:temps}), we calibrate to the absolute magnitude and temperature of the Sun \citep{Casagrande18}, giving a radius of $R_{2} = 0.27 \pm 0.02\,R_{\odot}$. Looking ahead to our interpretation of the source, this radius is $\sim5\times$ larger than white dwarf radii with similar colors/temperatures, explaining its relatively high optical luminosity (Figure~\ref{fig:cmd_fig}).

As an initial estimate of the mass of the secondary, we use our spectroscopic measurement of log $g = 4.6\pm0.2$ cm s$^{-2}$, which formally gives $M_2 = 0.11^{+0.06}_{-0.04} M_{\odot}$ when combined with the $R_2$ measurement above. While we do not take this precise value too seriously given the systematic uncertainties in the spectroscopic fit, the binary evolution modelling presented in Section~\ref{sec:MESA_models} predicts a secondary mass near the upper end of this range, suggesting a consistent scenario.

For a neutron star in the mass range $1.4-2.0\,M_{\odot}$, the implied mass ratio $q = M_2/M_1 \sim 0.03$--0.12, giving an effective Roche lobe radius in the range $\sim 0.6$--0.8 $R_{\odot}$ \citep{Eggleton83}. Hence, given $R_{2} = 0.27 \pm 0.02\,R_{\odot}$, the secondary in J1120 is strongly underfilling its Roche lobe ($R_2/R_L \lesssim 0.45$). A more massive secondary would have a larger Roche lobe and hence would underfill it to an even greater degree.

\begin{figure}[t!]
\begin{center}
	\includegraphics[width=\linewidth]{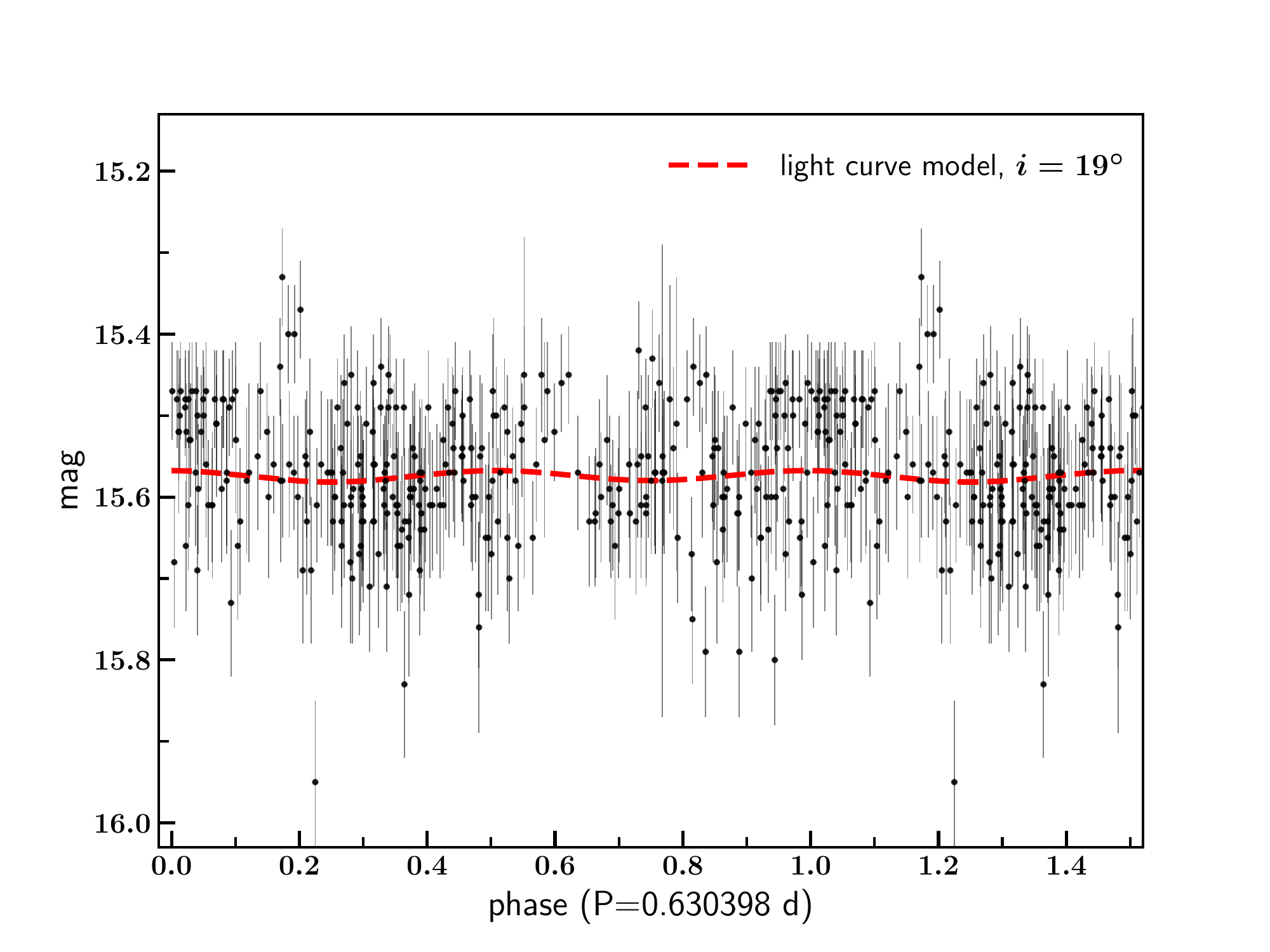}
    \caption{Unfiltered Catalina Sky Survey photometry of J1120 folded on our spectroscopic period, along with a light curve model assuming an inclination consistent with a neutron star-mass primary ($i=19^{\circ}$). The uncertainties on individual measurements are clearly much larger than the amplitude of the model light curve. Data have been repeated $1.5\times$.}
\label{fig:CSS_fig}
\end{center}
\vspace{-0.4cm}
\end{figure}

At the system inclinations inferred from Figure~\ref{fig:incl}, the amplitude of the optical light curve due to tidal distortion of the secondary would be $< 0.02$ mag in $V$-band \citep{Orosz00}, much less than the typical magnitude uncertainties from the ASAS-SN and CSS data (0.04 and 0.08, respectively). We show the CSS photometry folded on our spectroscopic period along with a light curve model assuming $i=19^{\circ}$ in Figure~\ref{fig:CSS_fig}.

Hence, the small size of the secondary relative to its Roche lobe and the face-on system inclination readily explain the lack of detectable orbital variability in the available optical data. With knowledge of the binary orbit from our spectroscopic results, future high-cadence, high-precision photometry could in principle be able to be searched for orbital modulation.

\subsection{High-Energy Properties}

Using the precise \emph{Gaia} parallax distance and the unabsorbed \emph{XMM} X-ray flux, the 0.5--10 keV X-ray luminosity is $(4.1\pm0.3) \times 10^{30}$ erg s$^{-1}$. The 0.1--100 GeV $\gamma$-ray luminosity is $L_{\gamma} = (1.35 \pm 0.05) \times 10^{33}$ erg s$^{-1}$, giving an X-ray to $\gamma$-ray luminosity ratio of $0.003$. This is lower than typical for redbacks \citep{Miller20}, with the only comparably low ratios observed for PSR J1622--0315 \citep{Gentile18,Strader19} and the unusual source PSR J1816+4510 \citep{Kaplan13}, discussed in more detail below. Instead, the X-ray to $\gamma$-ray flux ratios and X-ray luminosities are more similar to black widows (e.g., \citealt{Lee18}), though nearly all black widows have shorter orbital periods than J1120. Given the face-on inclination of the binary, any relativistic Doppler boosting is expected to be minimal compared to a typical spider MSP intrabinary shock.

While the relatively soft X-ray photon index from the \emph{XMM} observation and low X-ray luminosity are consistent with a weak, black widow-like intrabinary shock, they could also be consistent with thermal and/or magnetospheric emission entirely from the millisecond pulsar, with little or no contribution from a shock (e.g., \citealt{Bogdanov06}). Given the weaker wind expected from a warm, non-Roche lobe-filling companion compared to a cooler, more bloated star, and the marginal predicted irradiation of the secondary (the intrinsic optical surface flux is a factor of $\gtrsim 150$ larger than the $\gamma$-ray flux), the lack of a shock may indeed be expected.

\section{Discussion}
\label{sec:discussion}

The observations presented thus far suggest the secondary has unusual properties compared to the companions of most known MSPs associated with \emph{Fermi} sources. However, here we show that in the context of the canonical formation scenario for MSPs, this system actually appears to be on the standard evolutionary pathway for forming a MSP--He white dwarf binary, albeit one with an especially short orbital period and extremely low-mass white dwarf.

In this standard scenario, the secondary is born with a relatively low mass ($\lesssim2\,M_{\odot}$), and depending on the initial orbital period, can initiate mass transfer on the main sequence or as an evolved star. The accretion fully recycles the neutron star and the envelope of the secondary is completely stripped, leaving behind an extremely-low-mass ($\sim0.1$--0.3 $M_{\odot}$) Helium white dwarf companion \citep[an ELM white dwarf;][]{Brown10}.
There is strong theoretical and observational support for this scenario (e.g., \citealt{Savonije87,Rappaport95,Tauris99,Podsiadlowski02,Istrate14,Strader19}), which predicts a correlation between the final orbital period of the binary and the mass of the He white dwarf.

In recent years, ELM white dwarfs have seen increasing study outside of the context of MSP binaries (e.g., \citealt{Brown10,Kilic11,Brown20}), where surveys such as SDSS have allowed the selection of hot, high surface gravity objects with masses too low to have been formed except through close binary evolution. \emph{Gaia} has allowed the efficient expansion of these studies to include somewhat cooler and more bloated stars (e.g., \citealt{Pelisoli19,elBadry21}) which can be interpreted as ``pre-ELM" stars  \citep[e.g.,][]{Maxted14}, i.e., stripped stars still in the process of contracting to sit on a He white dwarf cooling track.

The properties of J1120 are fully consistent with being in this intermediate stage, on its way to eventually becoming a short period MSP--He white dwarf binary. We explore this hypothesis with binary evolution models in the following subsection.

\subsection{Binary Evolution Models: A pre-ELM companion}
\label{sec:MESA_models}
In order to understand the possible evolution of J1120, we performed a series of binary evolution simulations using Modules for Experiments in Stellar Astrophysics \citep[MESA, version 15140;][]{Paxton11,Paxton13,Paxton15,Paxton18,Paxton19}.

We initialize our models with a zero-age main sequence (ZAMS) star orbiting a more massive neutron star represented as a point mass. Mass transfer occurs once the companion overfills its Roche lobe with rates determined following the prescriptions of \citet{Kolb90} (\texttt{mdot\_scheme = `Kolb'}). The mass transfer (accretion) efficiency of the neutron star is defined as $\epsilon = 1 - (\alpha + \beta + \delta)$, where $\alpha$, $\beta$, and $\delta$ denote the mass fraction lost from the vicinity of the donor, accretor, and from a circumbinary coplanar toroid, respectively \citep{Tauris06}. For all our models we assume $\delta = 0$. We implement magnetic braking following the prescription from \citet{Rappaport83} with a braking index $\gamma_{\textrm{mb}} = 3$. The remainder of the MESA parameters were left to their default values.

Similar to the method employed by \citet{Bellm16} for their models of the redback binary PSR J2129--0429, we utilize our observational constraints on the companion orbital period, log $g$, radius, and temperature to establish feasible evolutionary models by requiring that these observables are matched at a particular age.

\begin{figure}[t]
\begin{center}
	\includegraphics[width=1.0\linewidth]{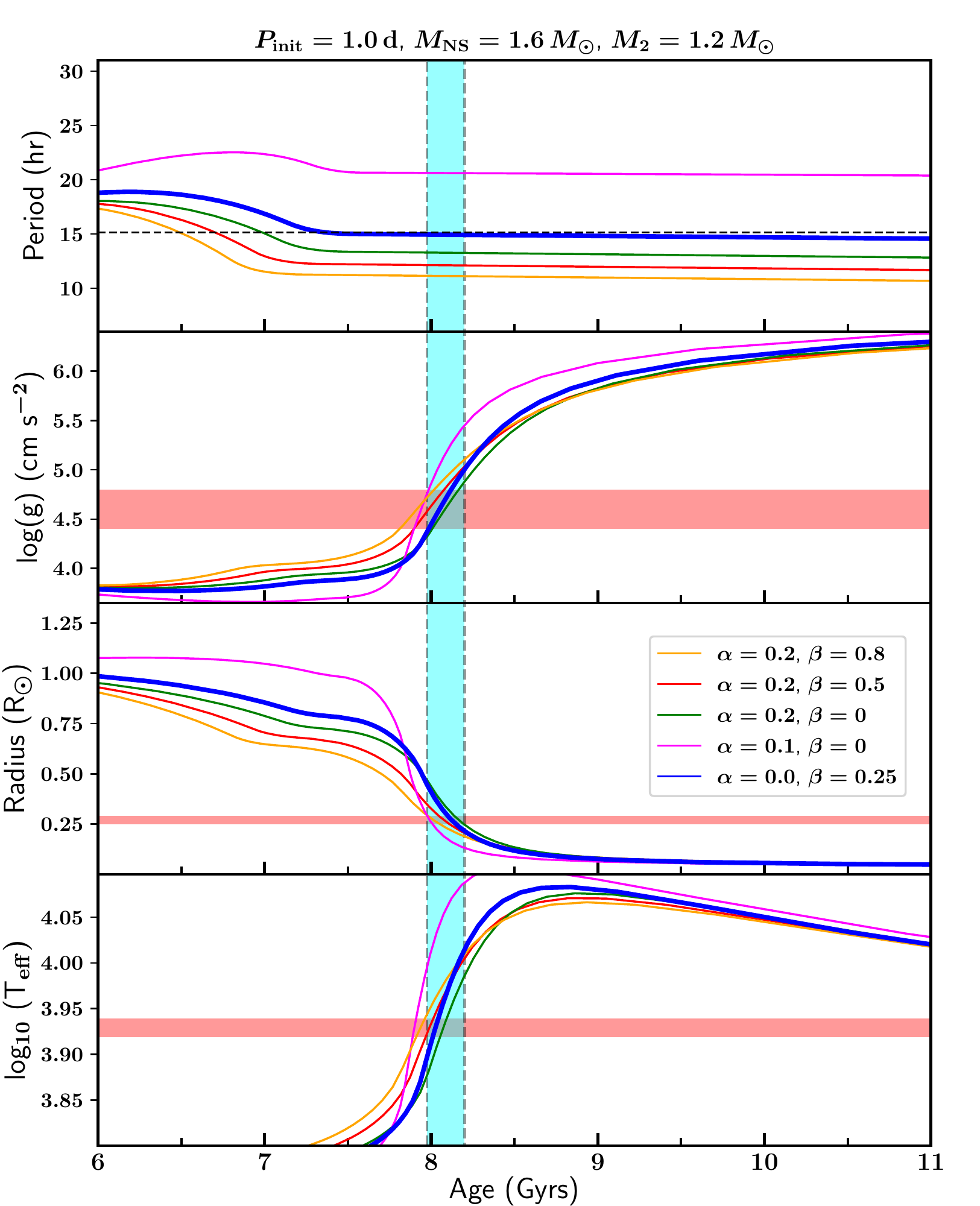}
    \caption{The evolution of the orbital period, log $g$, radius, and effective temperature of the companion for our MESA models simulating a $1.2\,M_{\odot}$ zero-age main sequence companion orbiting a $1.6\,M_{\odot}$ neutron star primary with an initial orbital period of 1.0 d. The different colors denote the effects of varying the mass loss efficiency at the companion ($\alpha$) and at the neutron star ($\beta$) (see text). In each panel, the horizontal shaded region corresponds to the range of values inferred from our observations. The ``best'' model is shown with a thick blue line, where all the model values are consistent with the observations at a single age (vertical shaded region).}
\label{fig:evolution_fig}
\end{center}
\end{figure}

We simulated a large range of models assuming a neutron star born with a mass of either 1.4 $M_{\odot}$ or 1.6 $M_{\odot}$, with initial orbital periods in the range 0.9--2.7 days and companion masses between 0.9--1.3 $M_{\odot}$. We show some of these models in Figure~\ref{fig:evolution_fig}, highlighting the effects of varying the accretion efficiency parameters $\alpha$ and $\beta$ (colored lines).

We note that the bifurcation period (separating models that diverge to long periods and those that evolve to shorter periods) is highly sensitive to the companion mass and the magnetic braking prescription. \citet{Soethe21} found that using a magnetic braking prescription that considers the wind mass loss and rotation of the donor as well as the magnetic fields generated by the outer convective zone \citep{Van19} shifts the bifurcation period to longer initial orbital periods. 

We first attempted to find compatible solutions assuming a relatively non-conservative mass transfer rate, 
$\alpha=0.2$ and $\beta=0.5$, implying $\epsilon=0.3$. We were unable to find models that reproduced our observations with this accretion efficiency, except for when $P_{\textrm{init}}=2.6$ d. However, the age at which this solution was compatible with our observations is $\sim$19 Gyr (greater than a Hubble time). At this $\epsilon$, models with smaller initial orbital periods were able to match the observed values for log $g$, $R_{2}$, and $T_{\rm eff}$ but none could simultaneously reproduce the observed period for J1120, always resulting in periods that were too low (e.g., the red model line in Figure~\ref{fig:evolution_fig}). A similar issue was found by \citet{Sanchez20} trying to model PSR J1012+5307, which also hosts a very low-mass He white dwarf.

For some models with shorter initial periods and larger mass transfer efficiencies, the results agree very well with our observations. The ``best'' model we find has $\epsilon \sim 0.75$, $P_{\rm init} = 1.0$ d, and initial component masses $M_{\rm NS} = 1.6\,M_{\odot}$ and $M_{2}=1.2\,M_{\odot}$. In Figure~\ref{fig:evolution_fig} we show this model in blue. With these initial conditions, mass transfer onto the neutron star begins while the secondary is still on the main sequence ($\sim$4 Gyr after ZAMS), and lasts for $\sim$3 Gyr. Figure~\ref{fig:evolution_fig} shows that at an age of $\sim$8.1 Gyr (vertical shaded region), all the model parameters match our observations (horizontal shaded regions). At this stage of the evolution, mass transfer is completely finished, as is the bulk of the orbital period evolution, and the companion is slowly contracting towards the He white dwarf cooling sequence, consistent with a pre-ELM companion.

The final component masses associated with this ``best'' model are $M_{\rm NS}\sim2.2\,M_{\odot}$ and $M_{2}=0.17\,M_{\odot}$. Although the final mass of the neutron star is somewhat high compared to known systems, due primarily to the low mass-loss efficiency, it is a generic prediction of evolutionary models that on average the binaries with shorter initial orbital periods will have more massive neutron stars \citep{Tauris99}. In addition, such neutron stars should have among the fastest spin periods and hence high spin-down luminosities, leading to a bias for \emph{Fermi} to find the fastest spinning (and more massive) MSPs. We note that this particular system does not appear to be among the more luminous \emph{Fermi}-detected MSPs (with a 0.1--100 GeV luminosity of only $\sim 1.4 \times 10^{33}$ erg s$^{-1}$), which could be either intrinsic or due to geometric factors. In any case, we do expect the neutron star to be massive. The neutron star mass could potentially be measurable if it were detected as a radio or $\gamma$-ray pulsar and if an improved proto-white dwarf mass could be measured from optical spectroscopy. 

Independent of our specific choices made for the MESA models, we can use other published orbital period--secondary mass correlations to check the robustness of our conclusions. \citet{Istrate16} predict that at solar metallicity, an orbital period of 0.6304 d implies a secondary mass of $0.17 M_{\odot}$, exactly matching our result. Even for other metallicities, a mass below $\sim 0.15 M_{\odot}$ or above $\sim 0.2 M_{\odot}$ seems relatively unlikely. While this is consistent with our rough mass estimate from fitting the medium-resolution SOAR spectra, it would be beneficial to have an improved white dwarf mass estimate from properly modeled high-resolution spectroscopy.

While we found no other comparably good solution for another combination of parameters, we did not search the entirety of the parameter space for all combinations of donor and accretor mass or efficiency, nor for varying assumptions about the magnetic braking law (e.g., \citealt{Van19}). Hence we do not claim that this is the only solution, but rather that this general area of initial condition parameter space can lead to systems whose properties match that of J1120 at the present day. Other evolutionary scenarios are not ruled out. Future, more detailed modeling will be useful once the suspected pulsar is found and timed, either in radio or $\gamma$-rays. This system may be especially valuable for further study due to its precise parallax, allowing an unusually accurate determination of the luminosity of the source.

\subsection{Comparison to Other Systems and The End Fate of 4FGL J1120.0--2204}
\label{sec:endfate}
The results presented above suggest a unique system is associated with 4FGL J1120.0--2204---a MSP binary with a pre-ELM white dwarf companion in a short-period orbit. To our knowledge, there is no other similar system, though the expected mechanism for forming MSP--ELM white dwarf binaries can generally explain its evolution. Perhaps the most similar MSP binary is PSR J1816+4510 \citep{Kaplan12,Kaplan13}, which is an eclipsing MSP often classified as a redback but with an unusually hot ($T_{\rm eff} \sim 16000$ K) secondary and a strongly supersolar metal abundance. While its mass and radius are inferred to be similar to that of J1120, it has a much shorter orbital period (8.7 hr), and the secondary may still be eroding in mass, more akin to known spider systems.

The end fate of J1120, after the secondary has contracted to the white dwarf cooling sequence, is a reasonable match to several known MSP--He white dwarf binaries. The MSPs PSR J1012+5307, PSR J1906+0055, and PSR J2006+0148 all have ELM white dwarf companions, with masses well established from radio pulsar timing campaigns and/or optical radial velocity measurements \citep{Sanchez20,Stovall16,Deneva21}. These systems have orbital periods ranging from $14.5-15.6$ hours, and secondary masses between $\sim0.12-0.17\,M_{\odot}$. These values are consistent with the orbital period and suspected mass of J1120 from our spectroscopic results and our binary evolution models, though at present the radius of J1120 is a factor of $\sim 6$ larger than these end-stage binaries. Our MESA models predict that in $\sim 2$ Gyr, J1120 will have a radius and log $g$ fully consistent with these ELM white dwarfs ($R_{2}\sim0.04\,R_{\odot}$, $\textrm{log}\;g\sim6.5$).

The long lifespan of the system in this intermediate stage means that such systems should not be too uncommon, and that more could be discovered through follow-up of unassociated \emph{Fermi} sources. Unlike the case for typical spider MSP binaries, pre-ELM MSP binaries may show little or no optical variability and weak X-ray emission, so would be best identified through their blue colors.

\begin{deluxetable}{lr}
\tablecaption{Summary of Properties}
\tablehead{
\colhead{Parameter} &
\colhead{Value}
}
\startdata
Opt. R.A. (ICRS h:m:s) & 11:19:58.309\\
Opt. Dec. (ICRS $^\circ$:$\arcmin$:$\arcsec$) & --22:04:56.33\\
$G_{\rm{mag}}$ (mag) & $15.538 \pm 0.003$\\
Period (d) & $0.630398 \pm 0.000027$ \\
$K_{2}$ (km s$^{\rm {-1}}$)  & $83.3 \pm 1.7$\\
$T_{0}$ (d) & $2459307.116 \pm 0.031$\\
$v_{sys}$ (km s$^{\rm {-1}}$) & $10.4 \pm 1.3$\\
$i$ ($^{\circ}$) & $\sim16 - 19$\\
$T_{\rm{eff}}$ (K) & $8500 \pm 200$\\
log $g$ (cm s$^{\rm {-2}}$) & $4.6 \pm 0.2$\\
$R_{2}$ ($R_{\odot}$) & $0.27 \pm 0.02$\\
$M_{2}$ ($M_{\odot}$) & $\sim 0.17$\\
$L_{X}/L_{\gamma}$ & $0.0030 \pm 0.0003$\\
$\mu_{\alpha}$cos $\delta$ (mas yr$^{\rm{-1}}$) & $-33.442 \pm 0.033$\\
$\mu_{\delta}$ (mas yr$^{\rm{-1}}$) & $14.182 \pm 0.028$\\
$(U,V,W)$ (km s$^{\rm {-1}}$) & ($130 \pm 4$, $-14.2 \pm 0.6$, $8.0 \pm 0.4$)\\
Distance (pc) & $816.8^{+22.0}_{-24.8}$\\[-10pt]
\label{tab:summary_table}
\enddata
\end{deluxetable}

\vspace{1cm}

\section{Conclusions and Future Work}
\label{sec:conclusions}

We have presented the discovery of the X-ray emitting compact binary inside the error ellipse of the unassociated \emph{Fermi} $\gamma$-ray source 4FGL J1120.0--2204. This source shows some properties similar to those of spider MSPs, including an overluminous secondary, a large spatial velocity, and a highly curved $\gamma$-ray spectrum, as well as other unusual properties such as an overabundance of Li. However, binary evolution modeling of the temperature, radius, gravity, and orbital period of the system together suggest the companion is a pre-ELM white dwarf star and that the binary is evolving toward being a short-period MSP--He white dwarf binary. We summarize many of the system properties in Table~\ref{tab:summary_table}.

In contrast to many recently discovered MSPs associated with \emph{Fermi} sources, which have tidally distorted or significantly irradiated secondaries, the optical companion shows no evidence for periodic orbital modulation in the optical light curves. This can be explained by a combination of the companion significantly underfilling its Roche lobe, a face-on orientation, and the rather low X-ray luminosity (implying a weak or absent intrabinary shock). The lack of orbital variability in this system highlights the value of using a range of approaches to identify counterparts to \emph{Fermi} sources.

For a neutron star primary, we find a face-on inclination of $\sim16^{\circ}$--$19^{\circ}$.
The a priori probability of such a face-on orientation is low, $\lesssim 6$\%, though there may be a selection bias: if the orientation were more edge-on, there might have been optical variability that would have allowed an earlier discovery of the binary. 

The X-ray luminosity of J1120 is somewhat low compared to most redbacks, though more similar to black widows. Together with the lack of orbital variability and the absence of optical emission lines, we find that it is plausible that there is no intrabinary shock in this system, with the X-ray emission entirely originating from the MSP.

Although J1120 appears to be on an evolutionary path that will eventually result in a binary that looks similar to several known short-period MSP--He white dwarfs binaries, there is no other known system in a similar pre-ELM stage. The best comparison is the strange eclipsing MSP PSR J1816+4510 \citep{Kaplan13}, which differs in several key aspects from 4FGL J1120.0--2204, but clearly deserves further attention.

The suspected MSP nature of 4FGL J1120.0--2204 is likely given the current data, but detecting the pulsar in either radio or $\gamma$-rays is still needed. The distance to the source is well-constrained, so detection of the pulsar would immediately allow a determination of the efficiency of the pulsar at converting its spin-down power to $\gamma$-rays. Pulsar searches of the 1FGL and 2FGL regions were performed a number of times between 2009--2012 with the GBT, GMRT, Effelsburg, and Parkes radio telescopes, and one brief search of the 3FGL region was executed in late-2016 with Parkes, all resulting in no detections (Pulsar Search Consortium, private comm.). Unlike typical spider MSPs, it is unclear whether 4FGL J1120.0--2204 should show radio eclipses, and it is possible that the radio beam from this pulsar is never pointed at Earth. Nonetheless, additional radio searches that use our optical ephemerides for scheduling would be valuable, and we have recently begun a GBT program utilizing this strategy. An independent, promising route to a pulsar detection would be to search for $\gamma$-ray pulsations, with the computationally-expensive search \citep[e.g.,][]{Nieder20,Clark21} constrained via our optical orbital parameters. The detection of the pulsar, combined with new high-resolution optical spectroscopy, could also allow a mass measurement of the neutron star, which is predicted to be high. Confirmation of an unusual and luminous MSP binary would provide yet more evidence of how incomplete the MSP census is, even at close distances ($< 1$ kpc) to the Sun.

\vspace{3.2cm}

\section*{Acknowledgements}
Special thanks to E. Brown and K. El-Badry for assistance with MESA.

This research was performed while SJS held a NRC Research Associateship award at the Naval Research Laboratory. Work at the Naval Research Laboratory is supported by NASA DPR S-15633-Y.

We also acknowledge support from NSF grant AST-1714825 and the Packard Foundation.

Some material is based upon work supported by NASA under award number 80GSFC21M0002.

Based on observations obtained at the Southern Astrophysical Research (SOAR) telescope, which is a joint project of the Minist\'{e}rio da Ci\^{e}ncia, Tecnologia e Inova\c{c}\~{o}es (MCTI/LNA) do Brasil, the US National Science Foundation’s NOIRLab, the University of North Carolina at Chapel Hill (UNC), and Michigan State University (MSU).

This work was based on observations obtained with XMM--Newton, an ESA science mission with instruments and contributions directly funded by ESA Member States and NASA.

We acknowledge the use of public data from the Swift data archive.

This research has made use of data and software provided by the High Energy Astrophysics Science Archive Research Center (HEASARC), which is a service of the Astrophysics Science Division at NASA/GSFC and the High Energy Astrophysics Division of the Smithsonian Astrophysical Observatory.

\bibliography{report}

\setcounter{table}{0}
\renewcommand{\thetable}{A\arabic{table}}

\begin{deluxetable}{lrr}
\tablecaption{Modified Barycentric Radial
Velocities of 4FGL J1120.0--2204}

\tablehead{
\colhead{MBJD} &
\colhead{RV} & 
\colhead{unc.} \\
\colhead{(d)} &
\colhead{(km s$^{-1}$)} &
\colhead{(km s$^{-1}$)} 
}
\startdata
58580.0292230 & 79.1 & 10.2 \\
59263.2108944 & --43.1 & 8.8 \\
59263.2252614 & --33.7 & 8.4 \\
59265.1105358 & --24.8 & 6.5 \\
59265.1607660 & 5.0 & 6.4 \\
59265.1747837 & 14.9 & 7.3 \\
59275.0714854 & --75.2 & 7.6 \\
59275.0854903 & --70.4 & 8.8 \\
59275.1026843 & --72.5 & 7.8 \\
59275.1169088 & --73.2 & 8.6 \\
59275.1483191 & --52.4 & 7.7 \\
59277.0835264 & --29.8 & 9.7 \\
59277.0978731 & --20.1 & 8.9 \\
59277.1145712 & --3.0 & 10.8 \\
59277.1287851 & --9.4 & 6.9 \\
59277.1454170 & 14.7 & 6.8 \\
59277.1578828 & 13.9 & 7.4 \\
59306.0077810 & --77.5 & 5.0 \\
59306.0268962 & --70.2 & 5.2 \\
59306.9965064 & 85.6 & 5.4 \\
59307.0163584 & 64.4 & 5.3 \\
59307.0370772 & 57.1 & 5.4 \\
59307.0552111 & 35.4 & 5.2 \\
59345.9653099 & 89.0 & 6.3 \\
59345.9831531 & 92.5 & 5.5 \\
59348.0915650 & --4.2 & 6.7 \\
59348.1100869 & --20.0 & 6.1 \\
59348.1304595 & --30.0 & 9.8 \\
59364.0891693 & --29.8 & 12.2 \\
59364.1609283 & 10.3 & 15.4 \\
59374.0215399 & --68.3 & 10.4 \\
59374.0405526 & --72.1 & 8.2 \\
59374.1021232 & --68.8 & 9.6 \\
59375.0215555 & 86.1 & 7.4 \\
59375.0362274 & 80.4 & 9.6 \\
59375.0838375 & 63.3 & 9.5 \\
59375.1181126 & 43.0 & 6.2 \\
59375.1320182 & 44.9 & 7.6 \\
59410.9891252 & 94.1 & 8.3 \\
59411.0032804 & 74.9 & 6.5 \\
\enddata
\tablecomments{ All dates are Modified Barycentric Julian Dates (BJD--2400000.5) on the TDB system \citep{Eastman10}.}
\label{tab:soar_rv_tab}
\end{deluxetable}

\end{document}